\documentclass[preprint,prb,amsmath,amssymb,pra]{revtex4}

\usepackage[english]{babel}
\usepackage{setspace}
\usepackage{bm}
\usepackage{amsmath}
\usepackage{graphicx}
\usepackage{xcolor}
\usepackage{soul}
\usepackage{ulem}
\usepackage[colorlinks=true, allcolors=blue]{hyperref}
\begin{document}

\title{ Local heating variations and transient effects in the coupling of thermal radiation and non-Fourier heat transport}

\author{D. Becerril}
\affiliation{Istituto di Struttura della Materia, CNR, 00133 Rome, Italy}
\author{A. Camacho de la Rosa, R. Esquivel-Sirvent}
\affiliation{Instituto de Física, Universidad
Nacional Autónoma de México, Ciudad de México 01000,
Mexico}

\begin{abstract}
In this work, we study the thermalization between two bodies separated by a vacuum gap by coupling the non-Fourier behavior of the materials with the radiative heat transfer in the near-field. Unlike the diffusion-type temperature profile, in non-Fourier materials, the temperature behaves as a wave, changing the thermalization process.  Due to the temperature profile induced by the coupling with conduction, we show that the radiative heat flux exchanged between the two bodies differs from the Fourier case, and exhibits transient temperature effects at the onset of the thermalization process. These results have important implications in nanoscale thermal management, near-field solid-state cooling, and nanoscale energy conversion.  
\end{abstract}

\maketitle

\section{Introduction} 

Thermalization is the process by which a system reaches a state of thermal equilibrium with its surroundings. Conduction and radiation are two fundamental modes of heat transfer that play a crucial role in a wide range of applications ranging from thermal therapies \cite{Andreozzi2019} to heat management at the nanoscale \cite{stewart2022control,smoyer2019brief}. In some cases, these two modes can interact, leading to a phenomenon known as conduction-radiation coupling.  In the near-field, thermalization can occur on a much faster timescale than in the far-field due to the proximity of the two bodies \cite{pendry1999radiative}.  Radiative heat transfer (RHT) between two bodies separated by a gap at different temperatures has become a topic of interest in energy management at the micro and nanoscale \cite{lucchesi2021radiative}. For large separations, this phenomenon is well described by the Planck theory of black body radiation, which only considers propagating electromagnetic waves \cite{cuevas2018radiative}. However, at subwavelength separations between the bodies, known as the near-field regime, strongly localized evanescent waves become the main contributors to RHT \cite{pascale2023perspective}. Unlike the far-field case, the heat flux depends on the separation between the bodies.  Close to room temperature of $300\;$K, thermal radiation in the far-field is limited to $\approx 6\; \text{Wm}^{-2} \text{K}^{-1}$ by Planck's blackbody radiation law. When tens of nanometers separate surfaces, radiative thermal conductance can increase several orders of magnitude above the black body prediction \cite{PhysRevB.4.3303,kittel05,Dewilde}. 
 In the original theory describing radiative heat exchanges between two closely spaced bodies introduced by Polder and van Hove \cite{PhysRevB.4.3303}, there is no coupling between the heat carriers inside the materials and the electromagnetic energy transferred across the vacuum gap.   

This interplay between heat transport and radiative heat transfer in thermal relaxation has been explored recently \cite{PhysRevLett.125.224302,PhysRevB.104.L100305,PhysRevB.94.121410}. In these works, the Boltzmann transport equation achieves a general coupling of heat radiation and conduction for bodies of arbitrary size. This leads to study systems exhibiting diffusive or semi-ballistic behavior \cite{hoogeboom2015new}.  These authors focus mainly on planar structures comprised of polar materials. It was concluded that the radiative heat flux exchanged between parallel slabs at nanometric distances is reduced due to the change in the temperature profiles within each body.  

 This paper addresses the case when heat is exchanged between non-Fourier materials. Non-Fourier heat transport refers to heat transfer in which the temperature gradient is not directly proportional to the heat flux. The heat flux can exhibit non-local, non-linear, or memory-dependent behavior, and the temperature field can propagate at a finite speed \cite{de2022causality}.

 To mend this unphysical consequence, Cattaneo and Vernotte \cite{cattaneo1958forme,vernotte1958paradoxes,Spigler,OSTOJASTARZEWSKI2009807}, independently proposed inserting a delay time $\tau$ between applying the temperature gradient and the heat flux \cite{maillet2019review}.    One example of a material that exhibits non-Fourier behavior is a thin film of metal, such as gold or silver, which can have a much higher thermal conductivity than the surrounding air or substrate \cite{ma15124287}. When a laser pulse is focused on such a material \cite{JAUNICH20085511}, the temperature of the metal can rapidly increase, leading to non-equilibrium behavior and the onset of thermalization. Further applications of non-Fourier heat conduction can be found in describing the temperature profile dynamics in biological tissue \cite{camacho2021time,gupta2019non}. 

In this work, we will discuss the coupling between radiation and conduction between two closely separated solids, provided the conduction properties do not follow Fourier's law of heat conduction.  For the separation range considered, radiative energy exchange competes with respect to the conductive heat transport within the bodies \cite{PhysRevB.94.121410}. We do not consider the transition from radiation to conduction at sub-nanometer gaps \cite{Chiloyan2015}.

\section{Theoretical framework}

\begin{figure} 
\centering
\includegraphics[width=0.5\textwidth]{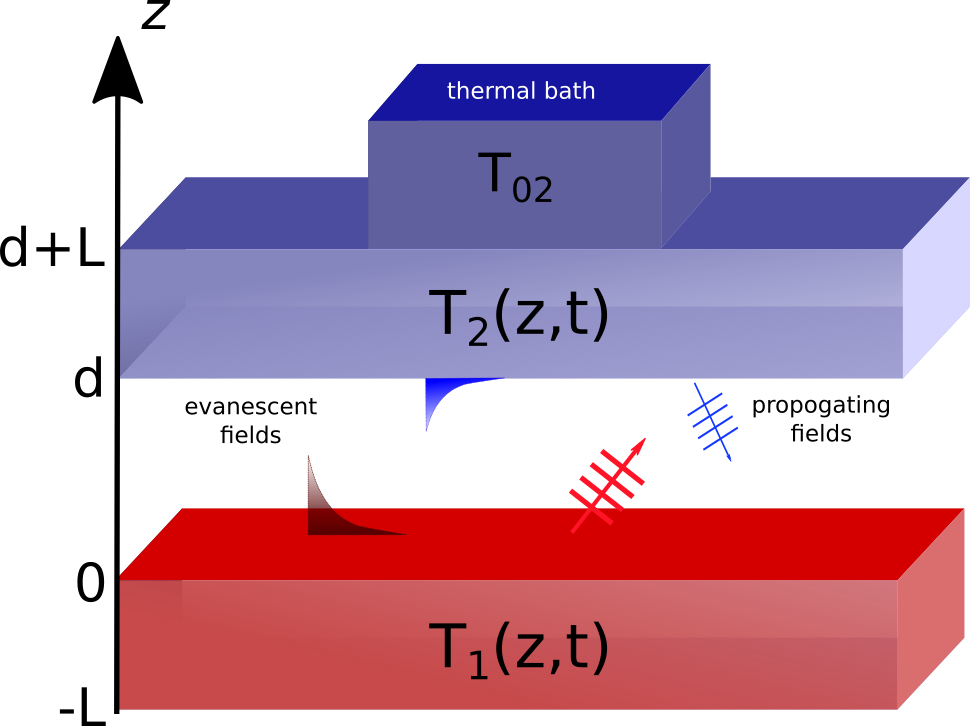}
\caption{Schematics of the  system under study. Slab 2 is in contact with a thermal bath at constant temperatures  $T_{20}$. A temperature profile is created within each slab.}\label{fig:system}
\end{figure}

We consider two planar bodies of width $L$ separated by a vacuum gap $d$ at initial temperatures $T_{01}$ and $T_{02}$ respectively, as shown in Fig.\ref{fig:system}. Body 2 is also connected to a thermal bath at temperature $T_{02}$. The bodies exchange energy only via RHT. Energy transferred via electromagnetic waves is then absorbed into the body and distributed according to a hyperbolic thermal transport equation, unlike Fourier materials that follow a parabolic-type equation. Within a slab, changes of the internal energy density $u(\vec{x},t)$  are due to the conductive and radiative energy fluxes, $\Phi_c$ and $\Phi_r$, thus
\begin{equation}\label{eq:energyconservation}
    \frac{\partial }{\partial t}u(\vec{x},t)= \Phi_c + \Phi_r.
\end{equation}

A constitutive equation can describe the conductive energy flux in both Fourier and non-Fourier materials. When the medium follows the Fourier equation, we will have $-\kappa \nabla T = \vec{q} $, where $\kappa$ is the thermal conductivity [W m$^{-1}$ K$^{-1}$], we assume it is a scalar valid for an isotropic medium. On the other hand, when the medium  is described by the Cattaneo-Vernotte equation, we have 
\begin{equation}\label{eq:cv1}
-\kappa \nabla  T = \vec{q} + \tau \frac{\partial \vec{q}}{\partial t}, 
\end{equation}
where $\tau$ is a delay time. The second term on the right-hand side can be seen as a first-order approximation to the description of the time lag between the heat flux $\vec{q}$ and changes in the temperature gradient $\nabla T$. For the limiting case of instantaneous response  $\tau\rightarrow 0$, the diffusive Fourier's law is recovered.

To obtain an equation for the temperature, we first notice that slab 2 is in contact with a thermal bath at constant temperature $T_{20}$, it will be useful to define the auxiliary function
\begin{equation}
F_k(\vec{x},t) = T_k(\vec{x},t) - T_{20},
\end{equation}
where $k=1$, 2 indicates the slab. The function $F_k$ describes deviations from the equilibrium temperature of the system and is therefore equal to zero when the slabs reach thermal equilibrium. To obtain a partial differential equation (PDE) for the temperature fields difference $F_k(\vec{x}, t)$, we can use the energy conservation equation Eq.\eqref{eq:energyconservation}, the CV heat conduction equation Eq.\eqref{eq:cv1}, and the constitutive equation for the internal energy $u = c_vT$, where $c_v$ is the specific heat per unit volume [J kg$^{-1}$ K$^{-1}$]. We get

\begin{equation}\label{eq:cv2}
\nabla^2 F_k(\vec{x},t) - \left(\frac{1}{\alpha} \frac{\partial }{\partial t} +\frac{1}{v^2}\frac{\partial^2}{\partial t^2} \right)F_k(\vec{x},t) = -\frac{1}{\kappa} \left( \Phi_r(x,t) + \tau \frac{\partial \Phi_r}{\partial t} \right),
\end{equation}
where $\alpha = \kappa /c_v$ is the thermal diffusivity [m$^2$ s$^{-1}$], $v^2 = \alpha/\tau$ is known as the heat propagation speed, and where $\Phi_r$ has taken the role of an external energy source.  Notice that while Eq.\eqref{eq:cv2} is the general 3-d expression, due to the symmetry of the problem, we only need to consider the spatial variable $z$, the axis perpendicular to the slabs. Eq.\eqref{eq:cv2} describes the time-dependent temperature profile within each slab. With no external energy source, Eq.\eqref{eq:cv2} describes a damped wave. This starkly contrasts to the Fourier case, where the temperature profile decays exponentially with time without an external energy source.


To obtain an expression of the radiative energy flux $\Phi_r$,  we further assume that the conductive channel is the dominant mode of energy transfer within each slab and therefore neglect RHT between portions of the same slab. Furthermore, we use the heat-sink approximation, valid for polar dielectrics, which considers RHT to occur only at the slab interface \cite{PhysRevB.104.L100305}. Application of this approximation allows us to neglect the source term $\Phi_r$ within each slab, and the PDE becomes homogeneous with initial and boundary conditions describing the RHT contribution of energy. We will, therefore, only require the description of $\left. \Phi_r\right|_{z = 0}$ and  $\left. \Phi_r\right|_{z = d}$ which will be imposed in the initial and boundary conditions.   

The radiative energy flux $\Phi_r$ evaluated at the interfaces can be related to the radiative thermal conductance defined as
\begin{equation}
    G_R = \lim_{\Delta T \rightarrow 0} \frac{\Phi^{1\rightarrow 2}(T_2+\Delta T,T_2)}{\Delta T},
\end{equation}
where $\Phi^{1\rightarrow2}$ is the net flux per unit area received by body 2 from 1.
Explicitly, the radiative thermal conductance can be written as

\begin{equation}
    G_R = \int_{0}^{\infty} \frac{d\omega}{2\pi}\frac{d\Theta(T,\omega)}{dT}\sum_{s,p} \int_{0}^{\infty} \frac{d\beta \beta}{2\pi}\left[\tau_\alpha^{prop}(\omega,\beta,d) + \tau_\alpha^{evan}(\omega,\beta,d)\right],
\end{equation}

\noindent
where $\beta$ is the magnitude of the wave vector component parallel to the surface, $\Theta(\omega,T) = \hbar\omega/[\exp(\hbar\omega/k_\beta T) - 1]$ is the Planckian function where $k_\beta$ is the Boltzmann constant and $\hbar$ is the Planck constant; $\tau_\alpha^{prop}$ and $\tau_\alpha^{evan}$, represent the P- and S-polarization transmission coefficient for propagating and evanescent waves respectively, which are given by
\begin{equation}
\tau_\alpha^{prop}(\omega,\beta,d) = \frac{(1-|r^\alpha|^2)^2}{|1-(r^\alpha )^2 \exp(2iQ_z d)|^2}
\end{equation}
for $\beta < \omega/c$, and 
\begin{equation}
\tau_\alpha^{evan}(\omega,\beta,d) = \frac{4[\text{Im}(r^\alpha)]^2\exp(-2|Q_z|d)}{|1-(r^\alpha)^2\exp(-2|Q_z|d)|^2}, 
\end{equation}
and in this case $r^\alpha$ are the usual Fresnel reflection coefficients, with $Q_z = \sqrt{(\omega/c)^2 - \beta^2}$ the z-component of the wave vector. The large increase in RHT at small separation distances is described by the term $\tau_\alpha^{evan}$.

As previously mentioned, instead of appearing explicitly in Eq.\eqref{eq:cv2}, the initial conditions hold the contribution of $\Phi_r$ at the interfaces. Since we have a second-order equation for $t$ and $z$, we require 2 initial and 2 boundary conditions. These can be choose by first observing that at an initial time, $t= 0$, each body has a constant temperature, $T_{01}$ and $T_{02}$, respectively so that  $F_1(z, t = 0) =  \Delta T$ and  $F_2(z,0) = 0$.  Notice that  since the bodies are at constant temperatures at the initial time $t=0$, there is no heat flux in each body $\Phi_c = 0 $, so at the initial time $t=0$, the energy conservation equation gives  $c_v\partial F_1(z,t)/\partial t\mid_{(t=0)} = \Phi_r = G_R \Delta T \delta(z)$ and $c_v\partial F_2(z,t)/\partial t\mid_{(t=0)} = \Phi_r =G_R \Delta T \delta(z-d)$. For  body 1 we further require that the heat flux be zero at the position farthest from the gap and for slab 2 we impose $F_2(z,t) = 0 $, since it is in contact with a thermal bath. Finally,
at the slab boundaries closest to the gap we require the conductive flux to be equal to the radiative flux.  The initial and boundary conditions can be summarized as
\begin{align*}
1) & \left. F_1(z,t) \right|_{t=0} = \Delta T, & 5) & \partial_z F_1(z = -L,t) =0             , \\
2) & \left.  F_2(z,t )\right|_{t=0} =0,         & 6) &  F_2(z = L+d,t) =0,   \\
3) & \left. \partial_t F_1(z,t ) \right|_{t=0} = -\frac{G_R}{c_v} \Delta T \delta(z),    & 7) &  \partial_z F_1(z = 0,t) =-\frac{G_R}{\kappa}\left[F_1(z=0,t) - F_2(d,t) \right], \; \text{and}   \\
4) & \left.\partial_t F_2(z,t ) \right|_{t=0} = \frac{G_R}{c_v}  \Delta T \delta(z-d),  & 8) &   \partial_z F_2(z = d,t) =-\frac{G_R}{\kappa}\left[F_1(z=0,t) - F_2(d,t) \right].     
\end{align*}

Notice that compared to the Fourier solution, in the CV case there is an additional initial condition for each slab, corresponding to conditions 3) and 4). The imposition of the initial and boundary conditions leads to a summation and is shown in the supplemental material. Solutions of temperature profiles can be written as
\begin{equation}\label{eq:CVsolution}
\begin{split}
T_1(z,t) &= T_{20} + \sum_{n=0}^{\infty}\left( a_n e^{r^+_{1n}t} + b_n e^{r^-_{1n}t} \right) \cos \left( x_n \frac{(z+L)}{L}\right) \\ 
T_2(z,t) &= T_{20} - \sum_{n=0}^{\infty}\left( a_n e^{r^{+}_{2n}t} + b_n e^{r^-_{2n}t} \right)\tan x_n \sin \left( x_n \frac{(z-L-d)}{L}\right),
\end{split}
\end{equation}

\noindent
with
\begin{equation}\label{eq:temporal_roots}
r_{in}^{\pm } = -\dfrac{v_i^2}{2\alpha_i }\left( 1 \mp \sqrt{1-4(x_n/L)^2 (\alpha_i/v_i)^2}\right),
\end{equation}
\noindent
where to find the constants $r^{\pm}_{in}$ and the series coefficients $(a_n,b_n,c_n,d_n)$,  we first have to solve the implicit equation that follows $x_n$, given by

\begin{equation}
x_n \tan (x_n) = \frac{2GL}{\kappa},
\end{equation}
which we then use to calculate these terms

$$u_n = \frac{L}{2}\left( 1+\frac{\sin(x_n)\cos(x_n)}{x_n}\right),$$ 
$$v_n = \frac{L}{x_n}\sin(x_n),$$
\begin{equation}
\begin{split}
     a_n  & =  -\frac{\Delta T}{r_n^+ - r_n^-}\left(r_n^- \frac{u_n}{v_n} -\frac{G_r }{L c_v}\frac{w_n}{v_n}  \right),   \\
     b_n & = \frac{\Delta T}{r_n^+ - r_n^-}\left( r_n^+ \frac{u_n}{v_n} - \frac{G_r }{L c_v}\frac{w_n}{v_n} \right),\\
   \text{and}\;  w_n & = \frac{\sin^2x_n \cos x_n}{4 x_n + \sin(4x_n)}.
\end{split}
\end{equation}
Notice that in our system, the material of slabs 1 and 2, are the same so that the subindex $i$ in Eq.\eqref{eq:temporal_roots} can be dropped. The thermalization dynamics described by the Fourier equation can be found following an analogous procedure and is  written here for the sake of completeness, a derivation of this expression can be found in the literature \cite{PhysRevB.104.L100305},
\begin{equation}\label{eq:FourierSol}
    \begin{split}
        T^F_1(z,t) & = T_{20} + 8\Delta T \sum_n w_n \cos x_n \cos\left[\frac{x_n(z+L)}{L}\right]\exp\left[-\frac{x_n^2 \alpha}{L^2}t\right], \\
                T^F_2(z,t) & = T_{20} -8 \Delta T \sum_n w_n (\cos x_n - 1)\cos\left[\frac{x_n(z+L)}{L}\right]\exp\left[-\frac{x_n^2 \alpha}{L^2}t\right]. \\
    \end{split}
\end{equation}

\section{Results and Discussion}

We begin by considering a system composed of two slabs of $L = 100$ $\mu$m composed of SiC separated by a vacuum gap of  $d = 1$ nm. As a first task, we show the deviations between the thermalization dynamics of the slabs when described by Fourier's law and the hyperbolic heat equation. A time delay $\tau$ must be chosen for the SiC slabs to do this. The time constant  for only a few materials has been reported \cite{Mitra95,Madhukar}, and a general theory to calculate the CV relaxation time constant for a given material is still lacking. Past work which describes hyperbolic thermal transport in semiconductor materials proposed using a time constant extrapolated from the phonon relaxation time. \cite{ZulKarnain2020}. While this approximation is reasonable, it is worth considering that the time constant $\tau$ in the CV describes the system heat flux time response to a thermal source, which is not necessarily the same as the phonon lifetime of the system. With this in mind, as a first approximation, we take  the delay time in the upper limit but within the range of known phonon lifetimes for SiC \cite{PhysRevLett.125.224302}.

\begin{figure}
\centering
\includegraphics[width=0.47\textwidth]{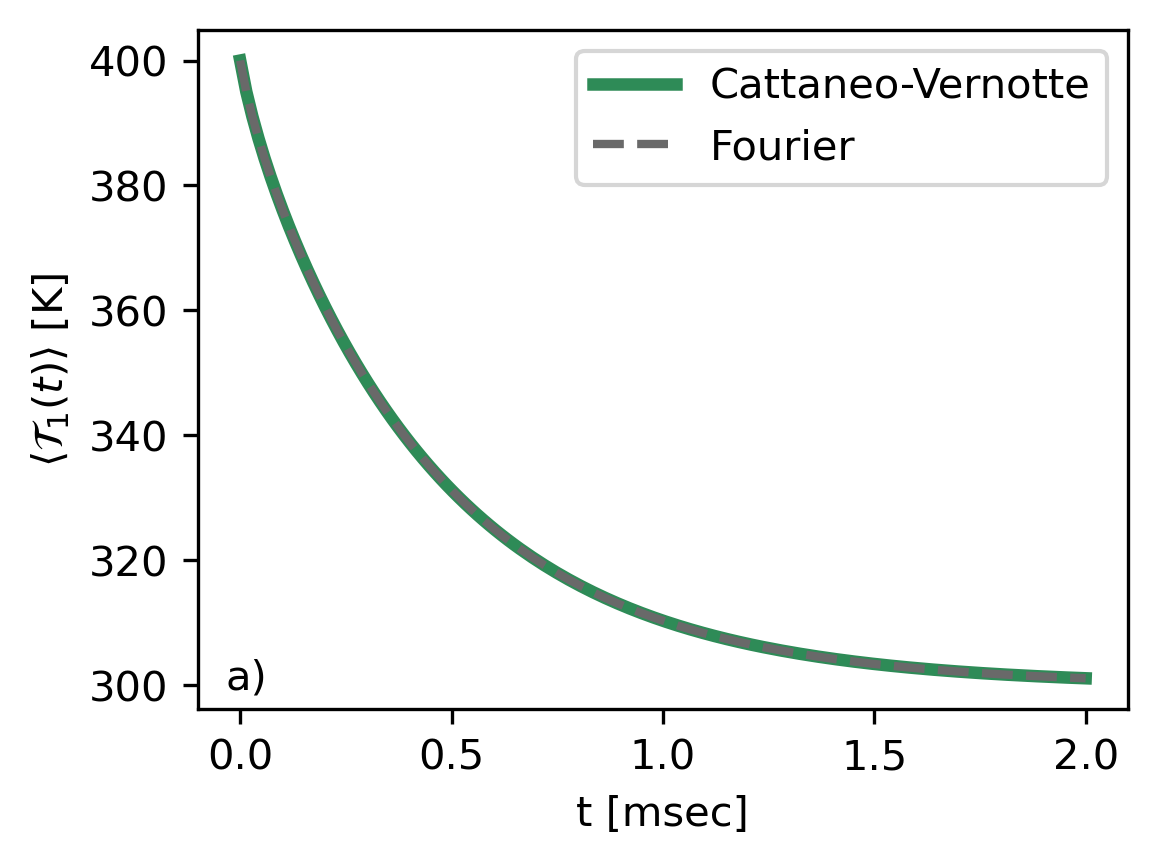}\includegraphics[width=0.53\textwidth]{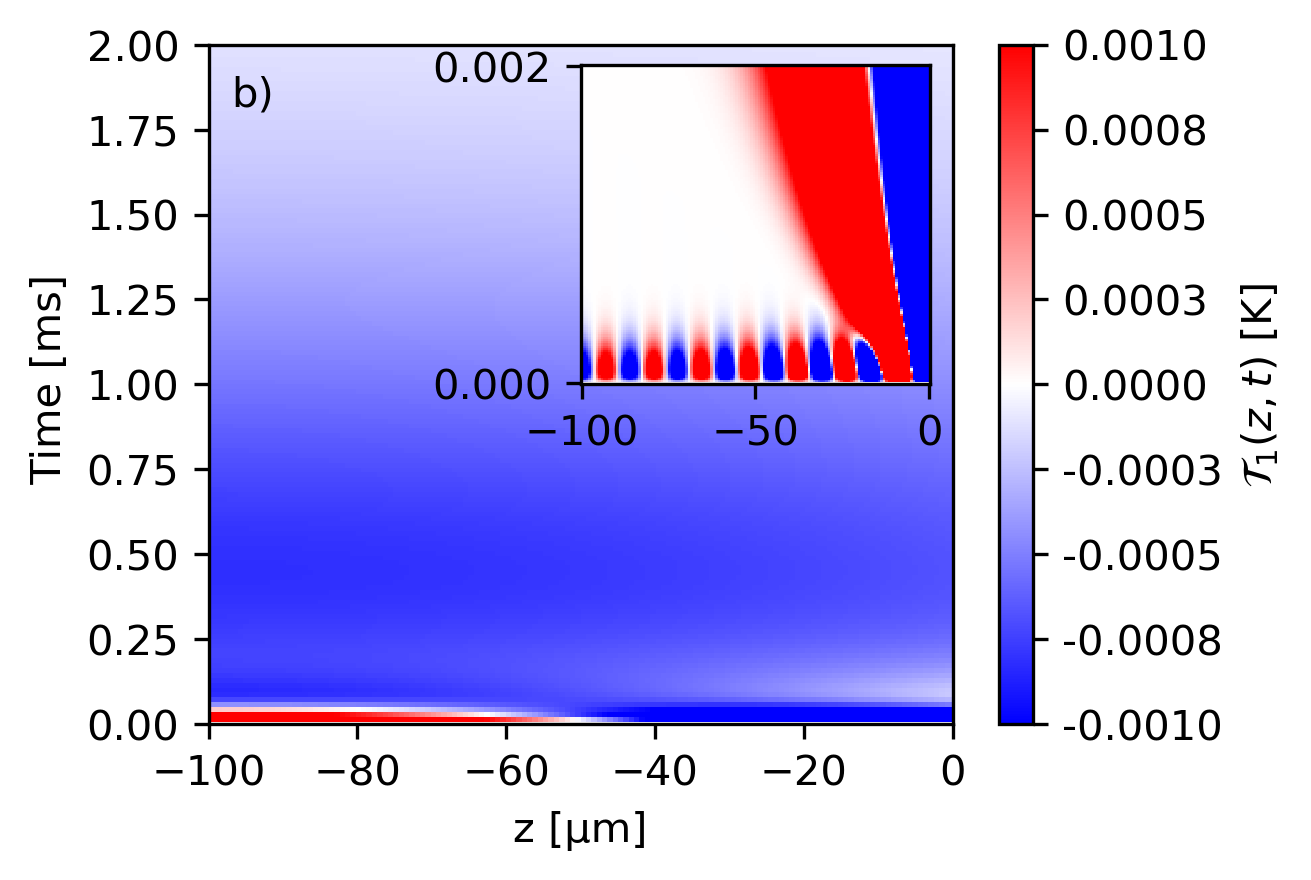}
\caption{(a) Time-dependence of the average spatial temperature in slab 1 using Fourier and CV equations. (b) Temperature difference between Fourier and CV equation $\mathcal{T}_1(z,t) = T^{CV}_{1}(z,t) - T_1^{F}(z,t)$. }\label{fig:results1}
\end{figure} 

In Fig.\ref{fig:results1}, we show the time-dependence of the spatial average temperature $\left<T_1(z,t)\right> = \frac{1}{L}\int_{-L}^0 T_1(z,t)dz$ of slab 1 when described by the Fourier and CV equation with a delay time $\tau = 10^{-8}$ s. Fig.\ref{fig:results1}(a) shows a nearly equal dynamic between the Fourier (dashed line) and hyperbolic (solid line) equations for such a small value of $\tau$, with the slabs reaching thermal equilibrium within the first few milliseconds in correspondence with past work \cite{PhysRevB.104.L100305}. In Fig.\ref{fig:results1}(b), we plot the temperature difference between the Fourier and CV case as a function of space and time within body 1, $\mathcal{T}_1(z,t) = T_1(z,t) - T_1^F(z,t)$. We show a zoom-in for $t<0.003$ ms $\approx 100 \; \tau$ in the upper right side, where we expect the largest wave-like behavior.
It can be seen that the temperature difference is small, with the difference tending to zero with increasing time, indicating the slab has reached thermal equilibrium. The inset of Fig.\ref{fig:results1}(b) shows a zoom of the first half millisecond, where we observe that $\mathcal{T}_1$ exhibits an oscillatory behavior that eventually dies out, indicative of a transient behavior in the thermalization process.
As expected, the difference between diffusive and hyperbolic behavior for small delay time values is negligible.
However, as the delay time grows, the difference between both conduction models is more noticeable. To further understand the origin of this dynamics difference, we point out that the spatial part of the solution is the same for Fourier and hyperbolic equations (Eq.\eqref{eq:CVsolution} and Eq.\eqref{eq:FourierSol}). The difference in the dynamics stems from the temporal part of the solution. 

\begin{figure}
\centering
\includegraphics[width=0.5\textwidth]{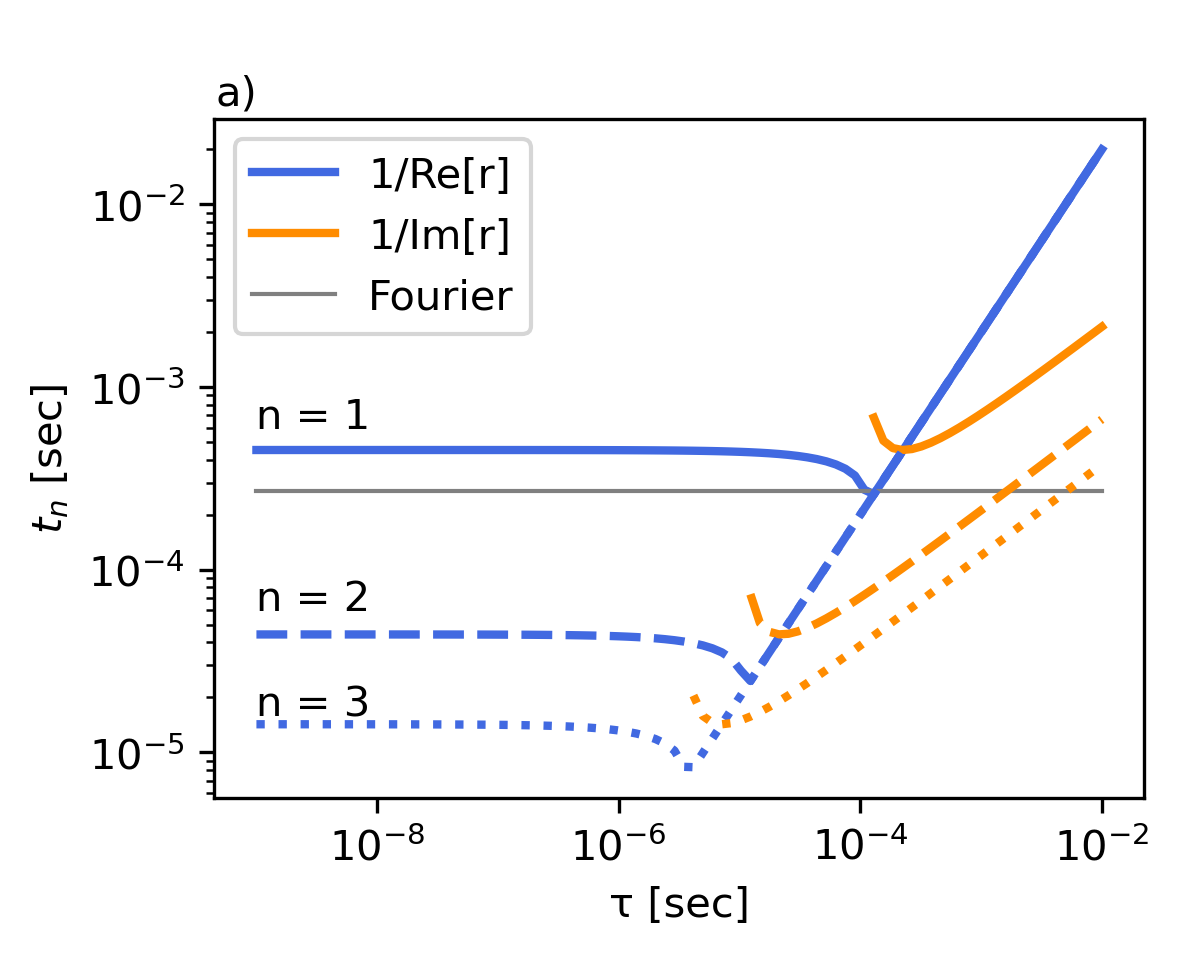}\includegraphics[width=0.5\textwidth]{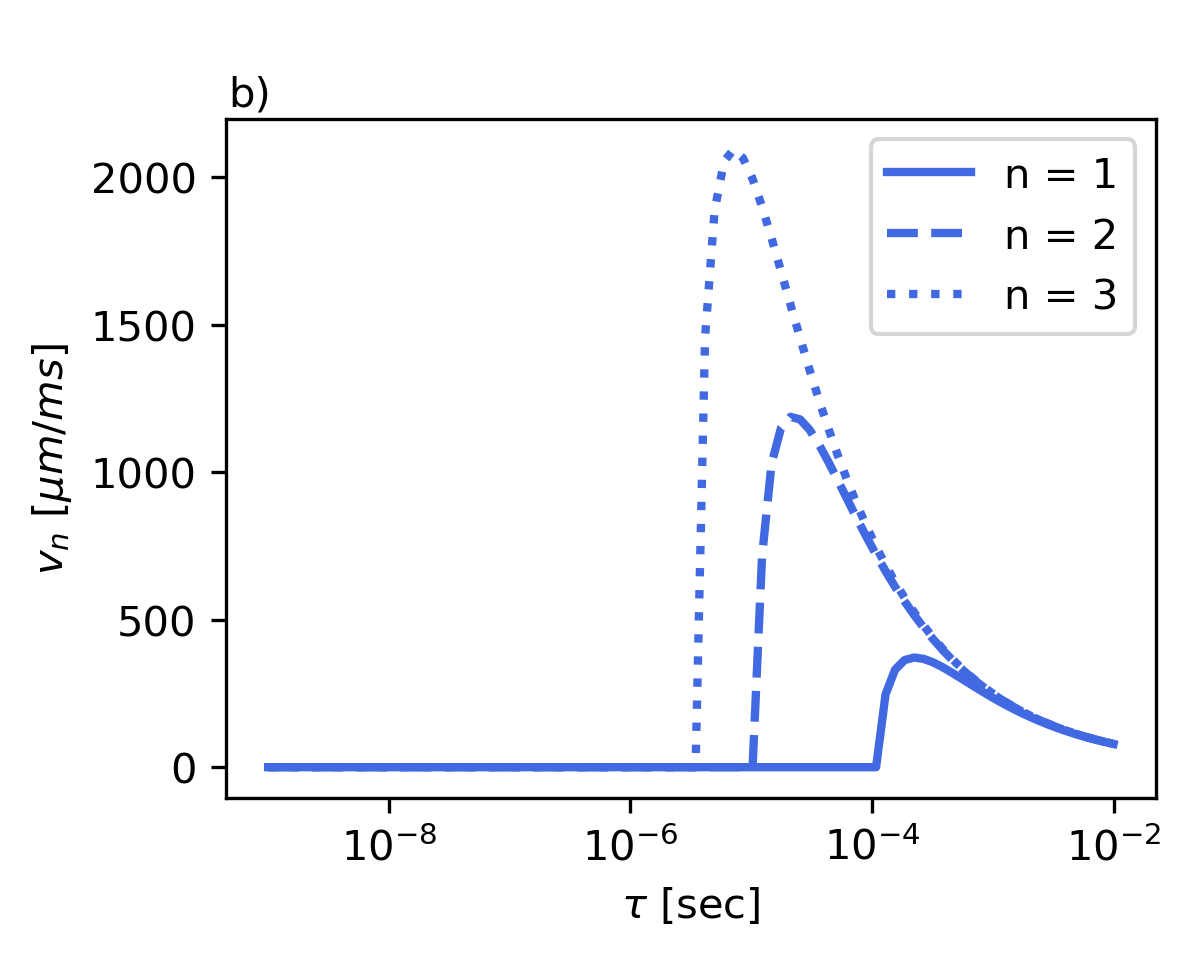}
\caption{(a) Lifetime of the first (solid) $n=1$ and second (dashed) $n=2$ modes as a function of the relaxation time $\tau$ for the system of SiC slabs and (b) mode velocity as a function of $\tau$.}\label{fig:lifetime}
\end{figure}

The roots that define the exponential behavior of the CV equation temporal solution are given by $r^{\pm}_{n}$, which are written in Eq.\eqref{eq:temporal_roots}. Unlike the Fourier case, with just one real root, when using the CV equation these roots can have an imaginary part depending on the system parameters. The dependence of the parameter $r^\pm_n$ as a function of the CV delay time $\tau$ is particularly interesting. The inverse of the real part of  coefficient $r^{\pm}_{n}$ defines a sort of lifetime of the mode $t_n = 1/r_{n}$, where $r_n=\text{min}(Re[r^+_n],Re[r^-_n])$, which can be compared to the analogous quantity obtained for the Fourier case, written as $1/r^F_n = L^2/(x_n^2 \alpha) $.
In Fig.\ref{fig:lifetime} we show the lifetimes of the first three modes $n = $ 1 (solid), 2 (dashed), and 3 (dotted) as a function of the CV response time $\tau$. 

The series of the temperature solution (Eq.\eqref{eq:CVsolution}) converges because as $n$ grows, the associated mode decreases its contribution. In Fig.\ref{fig:lifetime}(a) for each $\tau$ we observe that $t_1(\tau)\geq t_2(\tau) \geq t_3(\tau)$.  As we increase $\tau$, the imaginary part of $r_n$ takes relevance, indicating the threshold where is relevant the wave-like behavior. 
In contrast, the number of modes required to accurately describe the temperature dynamics near the temporal origin will depend on the value of $\tau$. As expected, the temporal window where a non-Fourier description is required depends on the value of $\tau$. For a value of $\tau = 10^{-8}$ s, modes with  $n \geq 2$ die off after less than a tenth of millisecond. In Fig.\ref{fig:lifetime}, we can see that, unlike the Fourier case, the temporal coefficient $r_n$ can have an imaginary part for a given $\tau$. It is interesting to note that for the case of $\tau = 10^{-8}$s, there is no imaginary part to the first mode. An imaginary part of $r_n$, and therefore the oscillatory nature of the solution, will be presented only for large values of $n$ with lifetimes of the order of $\tau$. 

To further illustrate the wave-like behavior of the solution, we point out that the physical temperature is obtained by taking the real part of Eq.\eqref{eq:CVsolution}. With this in mind, we can recast each term of the sum Eq.\eqref{eq:CVsolution} in the form $T(z,t) = T_{02} + \sum_{n = 1}^\infty\text{Re}[ T_{1n} + T_{2n}]$, where each term $T_{in}$ satisfies that
\begin{equation}
\begin{split}
\text{Re}[T_{in}]\propto & e^{-r_{in}^{\prime}t}\left\{ A_n^\prime \left[ \cos\left(r_{in}^{\prime\prime}t - \frac{x_n}{L}(z+L) \right) + \cos\left(r_n^{\prime\prime}t + \frac{x_n}{L}(z+L) \right)\right] \right. \\
 & \left. - A_n^{\prime\prime} \left[ \sin\left(r_{in}^{\prime\prime}t - \frac{x_n}{L}(z+L) \right) + \sin\left(r_{in}^{\prime\prime}t + \frac{x_n}{L}(z+L) \right)\right] \right\}, 
\end{split}
\end{equation}
where we write $r_{in} = r^\prime_{in} + i r^{\prime\prime}_{in}$, and the coefficient $A_n=A_n^\prime+iA_n^{\prime\prime}$, it saves the information of $a_n$ and $b_n$. This shows that the sum is just a superposition of damped waves. Furthermore, it allows us to define the velocity of each mode as $v_n = \frac{ r^{\prime\prime}_{i,n} L}{x_n} $. The velocity cannot be defined when no imaginary part of $r_i$ exists.  In Fig.\ref{fig:lifetime}, we show the velocity of the first three modes as a function of the parameter $\tau$. We can see that the mode velocity increases as $n$ increases, and even the slowest mode ($n=1$) travels 100 $\mu$m in about 0.2 milliseconds.
 
\begin{figure}
\centering
\includegraphics[width=0.5\textwidth]{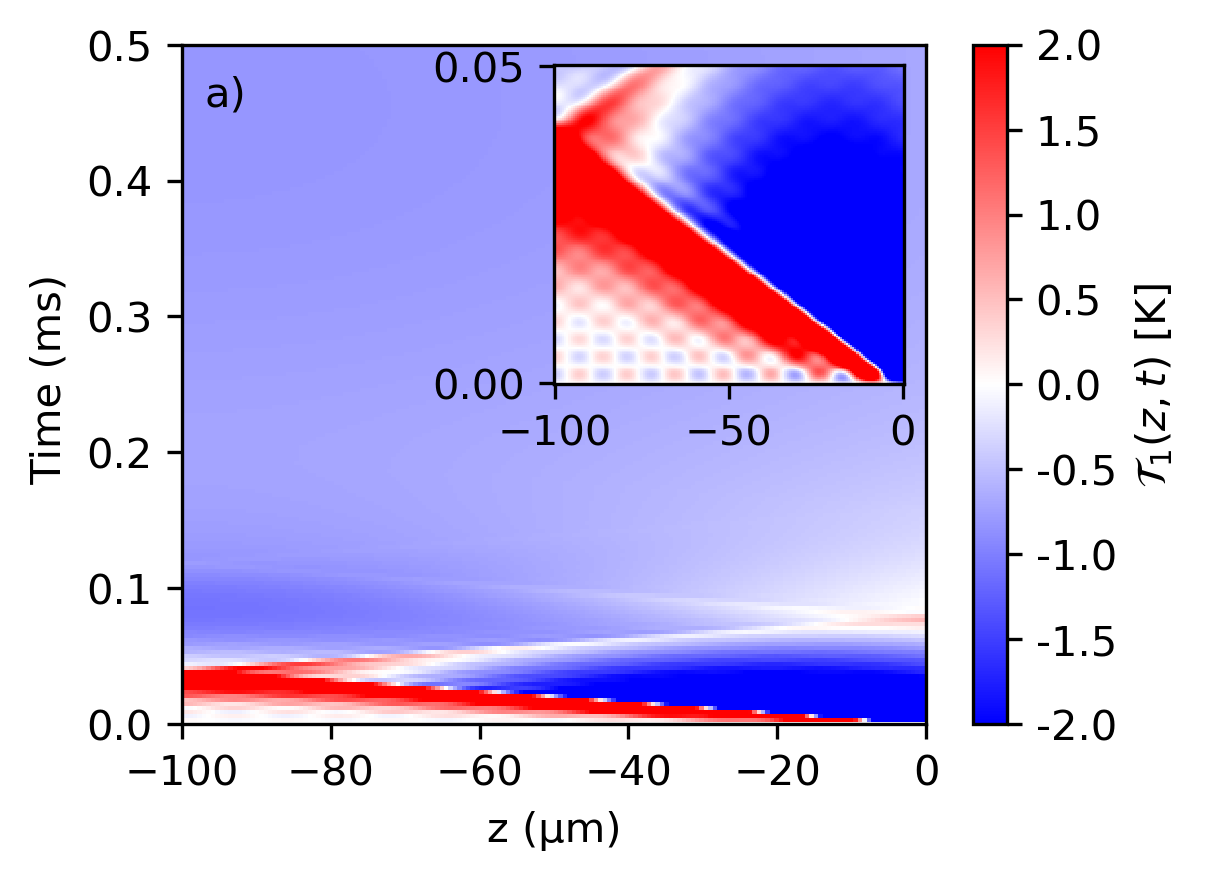}\includegraphics[width=0.5\textwidth]{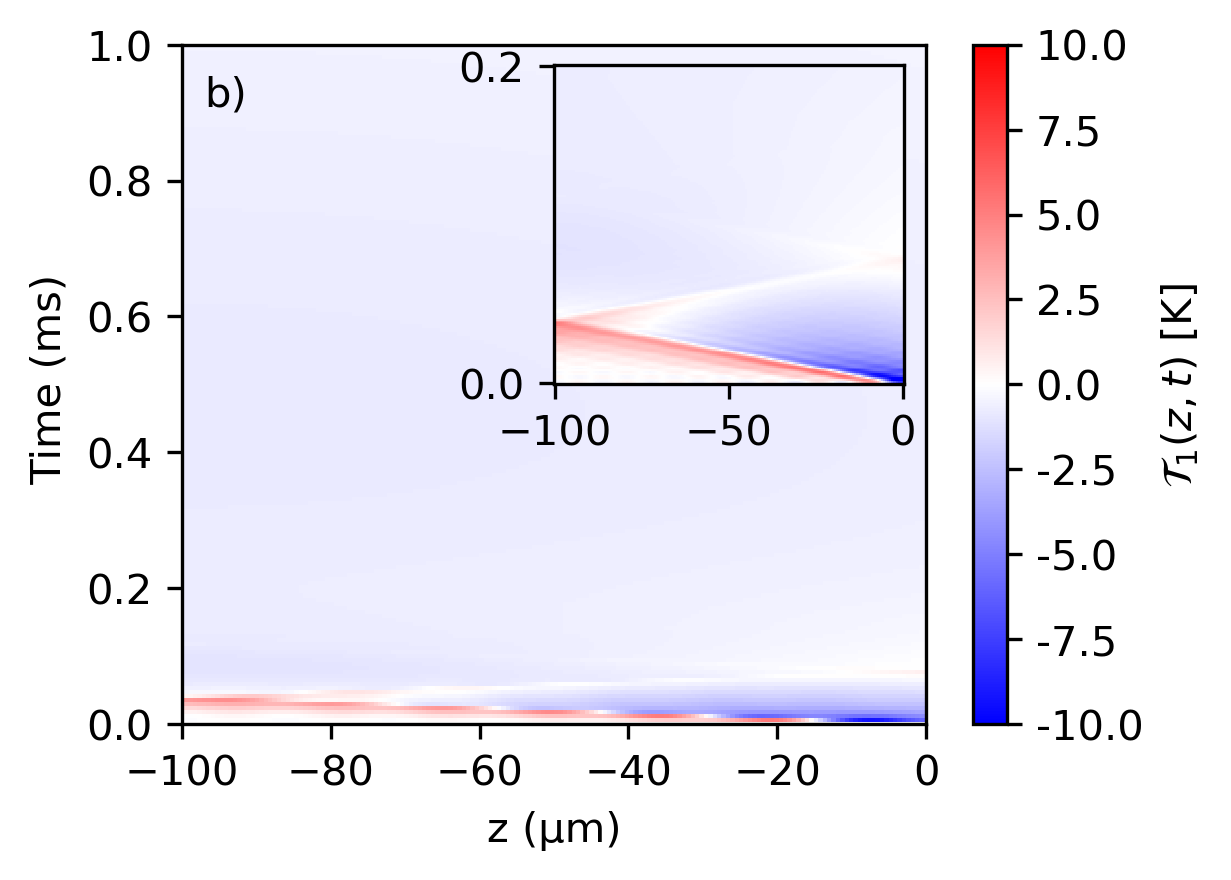}
\caption{Difference of the thermalization $\left< \mathcal{T}_1(z,t)\right>$ dynamics of slab 1 with (a) $\tau = 10^{-5}\;$s and (b) $\tau = 10^{-4}\;$s. }\label{fig:maptaus}
\end{figure}

While deviations from Fourier transport can be small when $\tau$ is of the order of the phonon lifetime, it is interesting to consider larger response times. Physically this can be obtained by inserting defects into the material so that the thermal response time no longer matches the  material phonon lifetime. To this end, we consider larger values of $\tau$, which may lead to oscillating modes with larger lifetimes $t_n$. In Fig.\ref{fig:maptaus}, we show the difference in thermalization dynamics for two different values of $\tau$. For these cases, the temperature differences are much larger than the corresponding small values of $\tau$, being of the order of tens of degrees. This larger deviation can be traced back to the coefficients $a_n(\tau)$ and $b_n(\tau)$ of Eq.\eqref{eq:CVsolution}, which depend on $\tau$ through parameter $r^\pm_n$. As in the previous case, the largest deviations are found after that onset the thermalization. The alternating differences in temperature shown in Fig.\ref{fig:maptaus} indicate the wave-like behavior of the CV solution.

We have established the influence of the RHT in the boundaries $z = 0$ and $z = d$, acting as energy sources. In the diffusive case, this energy perturbation is felt instantaneously in the whole slab,  while for the hyperbolic model, the introduction of $\tau$ implies a delay in the transport of energy; thus the farthest position felt later the thermal perturbation than the first point, giving rise to a wave-like behavior. In Fig.\ref{fig:maptaus}, we can see a minimum in $\mathcal{T}_1$ that reflects back and forth between the two interfaces of slab 1, it means that Cattaneo's model predicts an oscillation around Fourier’s law for the first moments, indicative of a transitory effect in the thermalization. As time increases, the hyperbolic solution predicts a lower temperature than the diffusive’s case, until both temperatures coincide and $\mathcal{T}_1 \rightarrow 0$. Both models reach the thermal equilibrium at similar times but with different rates of change.
For the case of $\tau =10^{-4}$ s in Fig.\ref{fig:maptaus}(b) a minimum in $\mathcal{T}_1$ is appreciable for nearly $\approx$ 1 ms = $10 \;\tau$ and travels at a speed roughly equal to  $v_1\approx 1000\; \mu $m/s, suggesting that this minimum is due mainly to the interference of mode n = 1 of the thermal waves (see Fig.\ref{fig:lifetime}(b)).

\begin{figure}
\centering
\includegraphics[width=0.54\textwidth]{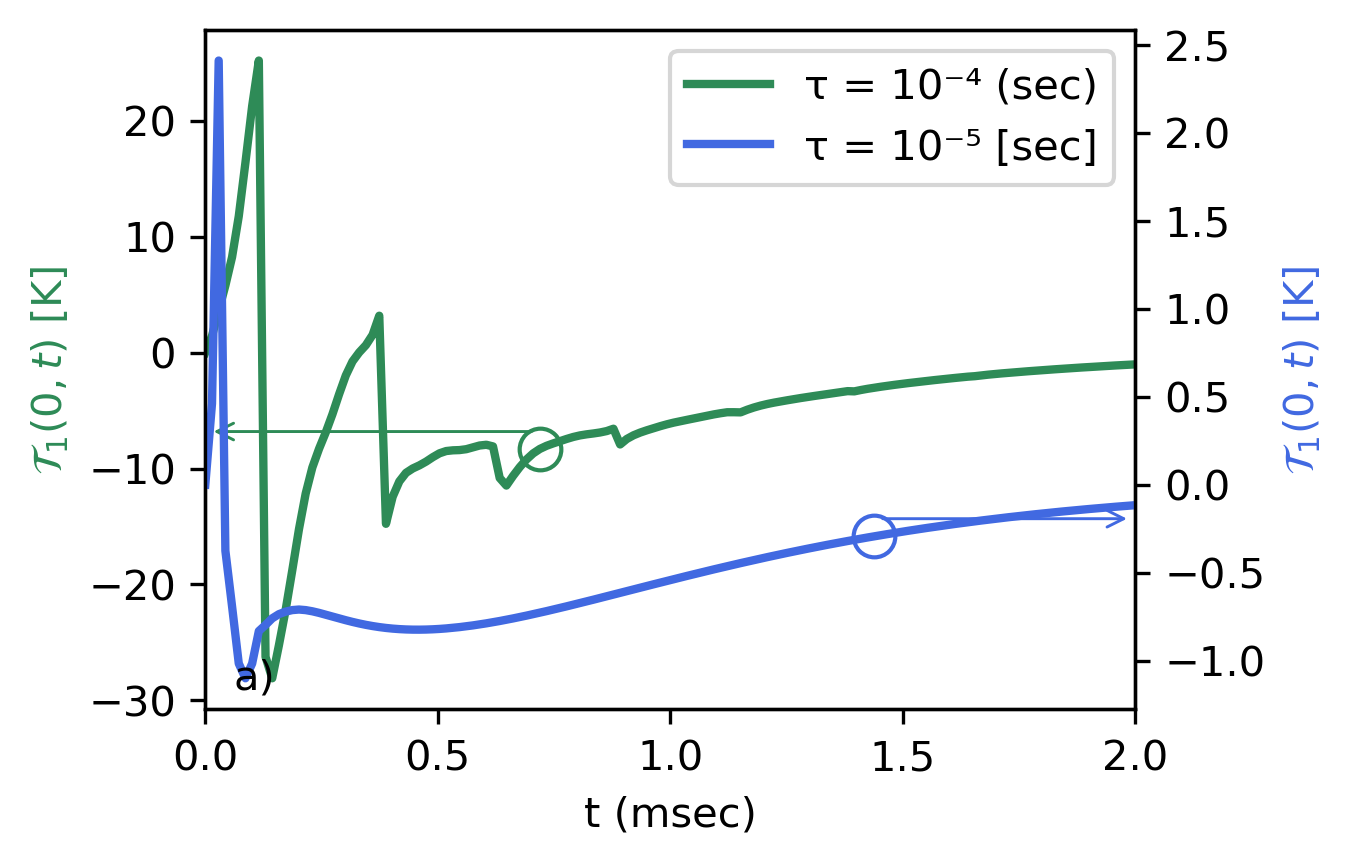}\includegraphics[width=0.46\textwidth]{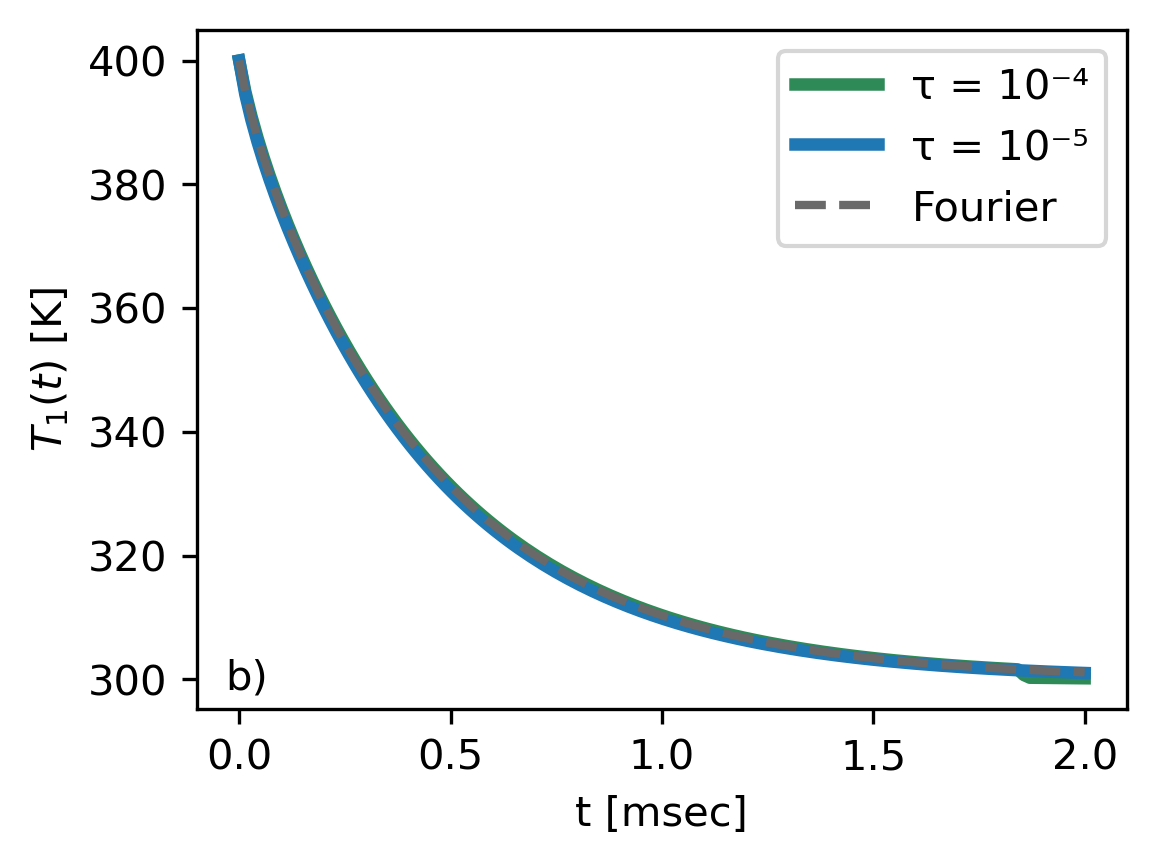}
\caption{(a) Difference of the thermalization $\mathcal{T}_1 = T_1(z,t) - T^F_{1}(z,t)$ dynamics evaluated at the interface of body 1 nearest to the gap $z = 0$ and (b) Averaged temperature within body 1 as a function of time $\left<T_{1}(z,t) \right>$ for $\tau = 10^{-5}\;$s and $\tau = 10^{-5}\;$s. }
\label{fig:mapcuts}
\end{figure}

In Fig.\ref{fig:mapcuts} we show the temperature difference evaluated at the interface of body 1 with the gap. This is equivalent to a vertical cut of the temperature maps of Fig.\ref{fig:maptaus} in $z=0$. In Fig.\ref{fig:mapcuts}(a) we can see that the oscillation periods are not constant during the thermalization, this is consistent with the $\mathcal{T}_1 = 0$ lines (isoclines) shown in Fig.\ref{fig:maptaus}. Besides, this agrees with the fact that at small times $t$, modes with larger $n$ and smaller periods contribute to the temperature profile. After half a millisecond the period becomes more regular since at this time, fewer modes have a perceptible contribution to the thermalization dynamic. In Fig.\ref{fig:mapcuts}(b), we show the averaged temperature within body 1, we observe as the diffusive and hyperbolic models reach the thermal equilibrium at similar times, but the rate of change of $\left<T_1(z,t) \right>$ is faster for $\tau=10^{-4}\;$s. It means that modifying the delay time affects the rate of change of the average temperature, decreasing faster for larger response times. While inside the body, the difference in the temperature profile can be up to around 40 degrees, the average temperature inside body 1 of Cattaneo's model for $\tau = 10^{-4}\;$s deviates only about 10 degrees with respect to the Fourier case. 

We observe that the solutions of the diffusive and hyperbolic models reach the thermal equilibrium in similar times, the dissimilarity between both models depends on the delay time. The two bodies exchange energy via near-field thermal radiation, and when coupled to heat conduction, it appears as a modification in the dynamics of the temperature profile. On the other hand, the diffusive model predicts a slower transitory process than the hyperbolic case for large values of $\tau$, however as time increases the difference between both models is negligible.

\section{Conclusions}
In this work, we have studied the thermalization dynamics of  two bodies that exchange energy via near-field thermal radiation using a hyperbolic heat conduction equation. For the case of SiC, we have concluded that if the thermal time response is of the order of the systems phonon lifetime, the thermalization dynamics described with the CV equation remains the same, to the order of a  millidegree, when compared to the Fourier description.  As the response time increases, the thermalization dynamics change, showing that the rate or speed of thermalization is slowed. Furthermore, at the onset of the thermalization process, the variation in temperature at the surface of the hotter body exhibits oscillations typical of transient behavior. These oscillations decrease with time, reaching a steady-state thermalization rate.  Large deviations in the temperature profile of tens of degrees can be expected when material response time is a tenth of a millisecond. This could have important implications in applications such as scanning probe lithography, where precise control of the temperature profile is required, particularly when considering temperature profiles in highly inhomogeneous material. Future work could also consider the case of layered systems of dielectrics and biological-like tissue, in particular, to have thermal shields with the potential of drastically slowing down the thermalization process. 

Finally, thermalization in the near-field is an important process in many physical systems and can be accurately described using the hyperbolic heat conduction equation. This equation considers the non-Fourier behavior of materials and allows for the rapid propagation of heat waves at velocities that exceed the speed of sound. Understanding thermalization in the near field is essential for designing microelectronic devices, optimizing laser-based manufacturing processes, and developing new materials with tailored thermal properties.

\acknowledgements{R.E-S acknowledges partial support from CONACyT grant A1-S-10537. D.B. acknowledges support from the Mexico City Ministry of Education, Science, Technology and Innovation (SECTEI) through grant CM-SECTEI/132/2022. }


%

\end{document}